\documentclass[twocolumn,showpacs,preprintnumbers]{revtex4}
\usepackage{graphicx}
\usepackage{dcolumn}
\usepackage{bm}

\begin{document}

\title{Variable Cosmological Constant model: the reconstruction equations
and constraints from current observational data}
\author{Yin-Zhe Ma}
\email{mayinzhe@itp.ac.cn}
\affiliation{Kavli Institute for Theoretical Physics China, Institute of Theoretical
Physics, Chinese Academy of Sciences (KITPC/ITP-CAS), P.O.Box 2735, Beijing
100080, People's Republic of China}

\begin{abstract}
In this paper we first give a brief review of the variable
cosmological constant model and its scalar field description. We
mainly discuss two types of variable cosmological constant models:
$a$ power law and $H$ power law models. A method to obtain all of
the equivalent scalar field potentials and the effective equation of
state of the two models is presented. In addition, the dynamics of
such scalar field potentials and effective equation of state are
discussed in detail. The parameters of the two models are
constrained by current 307 high-quality "Union" SN Ia data set,
baryon acoustic oscillation (BAO) measurement from the Sloan Digital
Sky Survey (SDSS), 9 observational \textrm{H(z)} data derived from
the Gemini Deep Deep Survey (GDDS) and the shift parameter of the
cosmic microwave background (CMB) given by the three-year Wilkinson
Microwave Anisotropy Probe (\textrm{WMAP}) observations. We also
calculate and draw the picture of the Hubble parameter, the
deceleration parameter and the matter density of the two models.
Then, we show that the indices $m$ and $n$ in the two models have
specific meaning in determining properties of the models. Moreover,
The reasons that indices $m$ and $n$ may also influence the behavior
of effective equation of state and scalar field potentials are
presented.
\end{abstract}

\maketitle


\section{Introduction}
In 1998, the discovery that the accelerated expansion of the Universe is
driven by the dark energy (DE) from the type Ia supernovae (SN Ia)
observations \cite{Riess98} greatly astonished the world. The Wilkinson
Microwave Anisotropy Probe \cite{Bennett03}, combined with more accurate SN
Ia data \cite{Riess04} indicates that the Universe is almost spatially flat
and the dark energy accounts for about 70\% of the total content of the
Universe. However, we know little about the nature of dark energy except for
its negative pressure. Therefore, a large number of works have been done in
recent years to explain this mystery.

The variable cosmological constant (hereafter VCC) \cite{Overduin98}
is one of the phenomenological ways to explain the dark energy
problem, because it is a straightforward modification of the
cosmological constant $\Lambda $ which enable itself to be
compatible with observations. Looking back to the history, we can
see that a lot of theorists have done numerous works to search for
the theoretical foundation of the VCC models and also investigate
the properties of the VCC models \cite{Strominger89}. In
\cite{Ozer86}, a model of $\Lambda \propto a^{-2}$ was proposed,
requiring that the cosmic density $\rho $ would equal to the
Einstein-de Sitter critical density $\rho _{c}$, which leads to a
closed Universe, without singularity, horizon, entropy and monopoly
problems \cite{Ozer86}. After that, it was also suggested a model
$\Lambda \propto a^{-2}$ ($\Lambda $ should be independent of $\hbar
$) with different initial conditions by \cite{Wu90}, which firstly
pointed out that time-dependent $\Lambda $ leads to the creation of
matter or radiation. Besides, a lot of work were done to propose
straightforward models relating $\Lambda $ to the Hubble parameter
$H(z)$: $\Lambda \propto H^{2}$
\cite{Freese87,Carvalho92,Lima93,Lima94}. Furthermore, people also
constructed a large number of phenomenal VCC models to describe the
dynamics of the Universe and there is a list in Ref.
\cite{Overduin98} summarizing the proposed models
\cite{Padmanabhan02}. There are also several papers
concerning the observational constraints about the VCC models \cite%
{Amendola06}.

In addition to the VCC models, scalar fields such as "quintessence" \cite%
{Peebles88}, "phantom" \cite{Caldwell99} and "quintom" \cite{Feng05} have
been introduced to effectively describe the dynamic dark energy, which are
distinguished by the effective equation of state (hereafter EEoS): $%
w_{DE}>-1 $, $w_{DE}<-1$ and $w_{DE}$ across $-1$ respectively. These models
are inspired by the fact that a decaying vacuum energy which has the very
high energy density at early time should be sufficiently small at present to
meet the current observation requirement, so they should evolve dynamically.
In order to obtain the corresponding quintessence potentials, the
reconstruction equations were derived and addressed the feasibility of the
approach by Monte-Carlo simulation \cite{Huterer98}; it was also constructed
the general scalar-field dark energy model \cite{Tsujikawa05} and developed
a method to construct them directly from EEoS function $w_{\phi }(z)$ \cite%
{Guo05,Lima03}. Moreover, some works have been done to reconstruct the
scalar potential from the scalar-tensor theory and investigate the modified
Newton theory \cite{Nojiri06}.

As a major part of our work, we analyze the EEoS and reconstruct the
potentials for two main types of VCC models----- the $a$ power law and the $%
H $ power law models----- from the point of view of dynamic scalar
fields. This work is necessary for people who are interested in the
coupled dark energy and dark matter \cite{Dalal01}, because such
models may avoid a lot of realistic problems such as the coincidence
problem. In addition, it is discussed how such phenomenological
models can be explained as a classical scalar field decaying into a
perfect fluid which might be interested by those who want to search
for the gravitational theory other than the general relativity,
because the Lagrangian in VCC should be different from the
Einstein-Hibert action in general relativity. Thus, this part should
be essential for people to see the possible forms of VCC models and
its corresponding scalar fields which are expected from string
theory or supergravity.

Another main part of this paper is to give an observational constraint on
the VCC models and explain the properties of cosmological parameters. This
part of work is the basic analysis to determine the right form of VCC models
from the observational requirement.

This paper is organized as follows: in section 2, we search for the
EEoS, the reconstruction equation and effective potentials for all
the $a$ power law models, generalizing previous work from Ref.
\cite{Maia01}. Next, we use the current observational data,
including 307 high-quality "Union" SN Ia data set, baryon acoustic
oscillation from SDSS, 9 observational \textrm{H(z)} data and CMB
shift parameter from \textrm{WMAP} three years result, to constrain
the index in this model. In addition, we analyze the properties of
the dark energy density, the dark matter density and the
deceleration parameter in this model. In section 3, parallelling to
section 2, we generalize the work from Ref. \cite{Maia01} and
discuss the EEoS, the reconstruction equation and reconstructed
potentials for all of the $H$ power law models. In addition, we also
give one example of this type of model to prove the effectiveness of
our method. Then, we use data to constrain the cosmological
parameters of this type of models and analyze the properties of the
dark matter density, Hubble parameter and the deceleration
parameter. The concluding remark will be presented in the last
section.
\section{$a$ power law models and corresponding potentials}
For a generalized VCC related to the scale factor $a$, we can write
\begin{equation}
\Lambda =Ba^{-m},  \label{Lamda for a}
\end{equation}%
where $B$ is a constant with the dimension of mass square and we call it the
dark energy amplitude. We assume the VCC is proportional to the scale factor
power $-m$, and power index $m$ plays a significant role in determining the
dark energy behavior as discussed below. Then, the dark energy density and
the Friedmann equation can be written as (Note that $a_{0}=1$, $\frac{d}{dt}%
=-H(1+z)\frac{d}{dz}$.)
\begin{equation}
\rho _{\Lambda }^{(a)}(z)=\ \Lambda M_{pl}^{2}=B(1+z)^{m}M_{pl}^{2}.
\label{rho Lamda for a}
\end{equation}%
\begin{equation}
3M_{pl}^{2}H^{2}=\rho _{m}^{(a)}+\rho _{\Lambda }^{(a)}.
\label{Friedmann for a}
\end{equation}%
For simplicity of calculations we assume spatial flatness $(k=0)$ which is
motivated by theoretical considerations, such as inflation, and also
confirmed by current observations such as \textrm{WMAP} three years result
\cite{Bennett03}. Our results can be easily generalized to the case with a
spatial curvature. We denote $M_{pl}=(8\pi G)^{-\frac{1}{2}}$ as the reduced
Planck mass and use superscript $(a)$ here to denote $a$ power law model, we
will also use superscript $(H)$ to denote $H$ power law model in the next
section. $\rho _{m}^{(a)}$is the dust matter density with the present value
\begin{equation}
\rho _{m0}^{(a)}=3H_{0}^{2}M_{pl}^{2}\Omega _{m0}.  \label{rho m0 for a}
\end{equation}%
Thus, we get
\begin{equation}
B=3H_{0}^{2}(1-\Omega _{m0}).  \label{B value}
\end{equation}%
In our practice, there is clearly only one degree of freedom in the $a$
power law model, which is the power index $m$. As for the VCC models, it is
rather natural to consider the interaction between the dark matter and dark
energy \cite{Chimento03}. Therefore, we should introduce an interacting term
$Q(z)$ with
\begin{equation}
\dot{\rho}_{m}^{(a)}+3H\rho _{m}^{(a)}=Q(z),  \label{Int matter (a)}
\end{equation}%
\begin{equation}
\dot{\rho}_{\Lambda }^{(a)}+3H(\rho _{\Lambda }^{(a)}+p_{\Lambda
}^{(a)})=-Q(z),  \label{Int energy (a)}
\end{equation}%
and the total energy conservation equation%
\begin{equation}
\dot{\rho}_{tot}+3H(\rho _{tot}+p_{tot})=0,  \label{energy total (a)}
\end{equation}%
still holds. Since VCC is the generalized from of cosmological constant, so
it satisfies $p_{\Lambda }^{(a)}=-\rho _{\Lambda }^{(a)}$. The Eq. (\ref{Int
energy (a)}) leads to
\begin{equation}
Q(z)=-\dot{\rho}_{\Lambda }^{(a)}=A(1+z)^{m}H,  \label{Q(z) for a}
\end{equation}%
where
\begin{equation}
A=3H_{0}^{2}(1-\Omega _{m0})mM_{pl}^{2},  \label{A value}
\end{equation}%
which means that the interaction is explicitly determined only by
the evolution of dark energy density.
\subsection{Interacting dark energy and reconstructed potentials in
the $a$ power law models} The anterior Eqs. (\ref{Int matter (a)})
and (\ref{Int energy (a)}) are the standard interacting dark energy
equations and the function $Q(z)$ represents the interaction between
dark energy and dark matter. Since the interaction may not be
directly observable, it is interesting to search for the
phenomenologically equivalent potentials which encode some
properties of the interaction.

One way to search for such a theory is to express the VCC models in
a field theory language, so the most straightforward way might be
the scalar field description. If one could find such a description
of the VCC models, it is natural to extend the scalar field
description to other space-time and other gravitational theory like
superstring theory. In addition, this description is very useful
since it provides a path to quantize the scalar field, which can
help people to understand the fundamental theory of the
phenomenological VCC models. Furthermore, the procedure to obtain a
scalar field description of a phenomenological model could be
applied to other models.

From this point of view, we want to see what are the EEoS and dark energy
potentials in the $a$ power law models. Changing the form of Eq. (\ref{Int
energy (a)}), we have
\begin{equation}
\dot{\rho}_{\Lambda }^{(a)}+3H(\rho _{\Lambda }^{(a)}+p_{\Lambda }^{(a)}+%
\frac{Q(z)}{3H})=0,  \label{energy eq for (a)}
\end{equation}%
so we could see the interaction $Q(z)$ contribute to the effective pressure
\begin{eqnarray}
p_{eff}^{(a)} &=&p_{\Lambda }^{(a)}+\frac{Q(z)}{3H}  \nonumber \\
&=&-\rho _{\Lambda }^{(a)}+\frac{Q(z)}{3H},  \label{p eff for a}
\end{eqnarray}%
so the EEoS of dark energy is
\begin{equation}
\omega _{eff}^{(a)}=\frac{p_{eff}^{(a)}(z)}{\rho _{\Lambda }^{(a)}(z)}=\frac{%
m}{3}-1,  \label{EoS-alaw}
\end{equation}%
so we obtain this result from the point of view of interacting dark
energy \cite{Lima03,Silveira94}. In Eq. (\ref{EoS-alaw}), the power
index $m$ is a constant, so the EEoS in the $a$ power law models are
all constants. We will see in the following subsection that the
best-fit of index $m$ constrained by current combined observational
data is $-0.09$, which means $\omega
_{eff}^{(a)}<-1$, so VCC is phantom-like \cite{Caldwell99}. However, for $%
2\sigma $ confidence level we cannot rule out the possibility that
$m>0$ (quintessence-like), so we should consider both the phantom
and quintessence scalar field potentials for the VCC model.

For a spatially homogeneous and isotropic scalar field, the effective energy
density $\rho _{\Lambda }^{(a)}$ and pressure $p_{eff}^{(a)}$ can be written
as
\begin{equation}
\mp \frac{1}{2}\dot{\phi}^{2}+V_{eff}^{(a)}(\phi )=\rho _{\Lambda }^{(a)},
\label{rho Lamda}
\end{equation}%
\begin{equation}
\mp \frac{1}{2}\dot{\phi}^{2}-V_{eff}^{(a)}(\phi )=p_{eff}^{(a)},
\label{p Lamda}
\end{equation}%
respectively, where upper (lower) sign represents the phantom (quintessence)
scalar field, and $V_{eff}^{(a)}(\phi )$ is the effective scalar field
potential for the $a$ power law models. At the same time, the effective
energy density $\rho _{\Lambda }^{(a)}$ and pressure $p_{eff}^{(a)}$ are
given by the interacting dark energy equations
\begin{equation}
\rho _{\Lambda }^{(a)}=A_{1}(1+z)^{m},  \label{rho Lamda1}
\end{equation}%
\begin{eqnarray}
p_{\Lambda }^{(a)} &=&-\rho _{\Lambda }^{(a)}+\frac{Q(z)}{3H}  \nonumber \\
&=&A_{2}(1+z)^{m},  \label{p Lamda1}
\end{eqnarray}%
where
\begin{equation}
A_{1}=3H_{0}^{2}M_{pl}^{2}(1-\Omega _{m0}),  \label{A1}
\end{equation}%
and
\begin{equation}
A_{2}=A_{1}(-1+\frac{m}{3}).  \label{A2}
\end{equation}%
We define the dimensionless quantities
\begin{equation}
\tilde{\phi}\equiv \phi /M_{pl}\text{, }\tilde{V}%
_{eff}=V_{eff}/3H_{0}^{2}M_{pl}^{2}.  \label{denote for phi}
\end{equation}%
Thus, the scalar field potential can be written as a function of redshift $z$
\begin{eqnarray}
\tilde{V}_{eff}^{(a)}(z) &=&\frac{1}{2}(\rho _{\Lambda }^{(a)}-p_{eff}^{(a)})
\nonumber \\
&=&A_{3}(1+z)^{m}.  \label{Veff for a}
\end{eqnarray}%
where
\begin{equation}
A_{3}=(1-\Omega _{m0})(1-\frac{m}{6}).  \label{A3}
\end{equation}%
Combining Eqs. (\ref{rho Lamda}) and (\ref{p Lamda}), we have
\begin{equation}
\frac{d\tilde{\phi}}{dz}=\mp \frac{C_{1}}{(1+z)[C_{2}+C_{3}(1+z)^{3-m}]^{%
\frac{1}{2}}},  \label{diff for phi}
\end{equation}%
where
\begin{equation}
C_{1}=(1-\Omega _{m0})^{\frac{1}{2}}\times \lbrack \left\vert m\right\vert
(3-m)]^{\frac{1}{2}},  \label{C1}
\end{equation}%
\begin{equation}
C_{2}=3(1-\Omega _{m0}),\text{ }C_{3}=(3\Omega _{m0}-m).  \label{C2}
\end{equation}%
The upper (lower) sign in Eq. (\ref{diff for phi}) represent $\dot{\phi}>0$ $%
(\dot{\phi}<0)$. In fact, the sigh is arbitrarily determined by assumption,
as it can be changed by $\phi \rightarrow -\phi .$ We choose the upper sign
in the following discussion. If we shift $\phi _{0}$ value, the potential in
the following figure will be shifted horizontally, but the shift doesn't
influence the whole shape of the potential. The field could be integrated
analytically as
\begin{equation}
\tilde{\phi}(z)=C_{4}\times \tanh ^{-1}[\frac{C_{2}^{\frac{1}{2}}}{%
(C_{2}+C_{3}(1+z)^{3-m})^{\frac{1}{2}}}],  \label{phi for a}
\end{equation}%
where
\begin{equation}
C_{4}=\frac{2}{\sqrt{3}}[\frac{\left\vert m\right\vert }{3-m}]^{\frac{1}{2}}.
\label{C4}
\end{equation}%
We let the integral constant equals to zero since the initial value
of field is meaningless. Solving this for $(1+z)$ and substituting
the result into (\ref{Veff for a}), we obtain the potential of the
$a$ power law model
\begin{equation}
\tilde{V}_{eff}^{(a)}(\tilde{\phi})=A_{3}[\frac{C_{2}}{C_{3}}\times (\coth
^{2}(\frac{\tilde{\phi}}{C_{4}})-1)]^{\frac{m}{3-m}}\text{.}
\label{alaw potential}
\end{equation}%
\begin{figure}[tbh]
\includegraphics[bb=0 0 347
237,width=3.4in,height=2.5in]{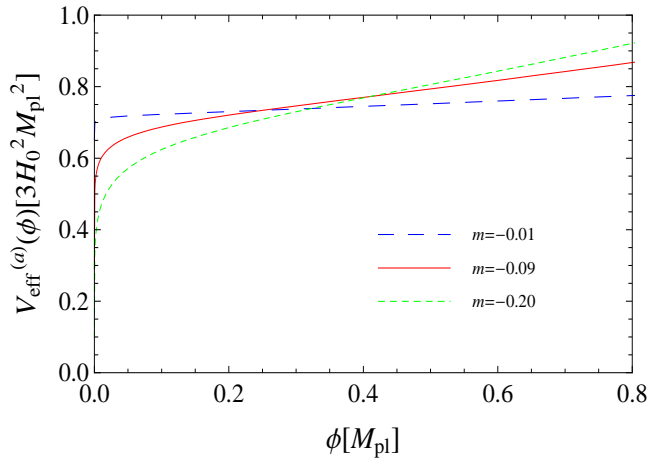}
\caption{Reconstructed potentials for $a$ power law model. Here we set $%
\Omega _{m0}=0.28$.}
\label{apotential}
\end{figure}
We use the best fits value for $m$ in the next section to draw the pictures
for the $a$ power law models' equivalent potentials. There are three main
characteristics for these potentials: First, they are all runaway type and
the whole shape doesn't change if $\phi $ is shifted horizontally. Second,
Eqs. (\ref{Veff for a}) and (\ref{phi for a}) determine that $\phi $
increases and $V_{eff}^{(a)}(\phi )$ increases as the redshift $z$ decreases
from large value to $-1$, which means that the dark energy potentials
increase as the Universe expands. From the figure, we could see that more
negative the value of $m$ is, the sharper the potentials will increase as
the field evolves. One of the possible explanation for this phenomenon is
that the more negative value of $m$ means the dark matter "decays" into dark
energy quicker, so the dark energy increase its potential value and energy
density faster. Third, the hyperbolic $coth$ function in the expression (\ref%
{alaw potential}) makes the $a$ power law potentials have the asymptotic
value. This is very interesting because one could obtain such behavior in
general in the supersymmetric QFT. This runaway form of potential is also
the one expected in the unstable D-brane system in superstring theory \cite%
{Sen05}.
\subsection{Hubble parameter and results of the constraints on $m$}
In this subsection, we want to constrain the parameter $m$ from
combined observational data, so we should obtain the Hubble
parameter and the luminosity
distance. We can change the variable $t$ to redshift $z$ in the Eq. (\ref%
{Int matter (a)}) to figure out the analytical expression for the matter
density
\begin{equation}
\rho _{m}^{(a)}(z;m)=3H_{0}^{2}M_{pl}^{2}[D_{1}(1+z)^{3}+D_{2}(1+z)^{m}],
\label{rho alaw}
\end{equation}%
where
\begin{equation}
D_{1}=\frac{C_{3}}{3-m},\text{ }D_{2}=\frac{m}{3-m}(1-\Omega _{m0}).
\label{D1}
\end{equation}%
We can easily note that the effect of VCC is just like a small perturbation
to the evolution of the matter density. If the evolution behavior of the
matter density doesn't deviate much from $(1+z)^{3}$ behavior ($\Lambda $%
CDM), the value of $m$ should be very near zero, which indicate that
even though the dark energy is not constant through the evolution of
the Universe, it at least should evolve very slow. This property
will be convinced through observational constraints on power index
$m$ in the following subsection. This equation is essential for our
purpose to solve the Hubble parameter in the following subsection.

As there is only one free parameter in this kind of power law models, it is
rather easy to obtain the best fit value from the current observational
data. We do this fitting using the high quality type Ia supernovae, baryon
acoustic oscillation from SDSS, observational $H(z)$ data and the the shift
parameter of the cosmic microwave background (CMB) given by \textrm{WMAP}
three years results. We discuss this problem in the framework of interacting
dark energy and accelerating Universe (see relevant work \cite{Silveira97}).

Integrating the Eq. (\ref{Friedmann for a}), we have the following equation
\begin{equation}
H(z)=H_{0}[D_{1}(1+z)^{3}+D_{3}(1+z)^{m}]^{\frac{1}{2}},  \label{Hubble alaw}
\end{equation}%
where
\begin{equation}
D_{3}=\frac{C_{2}}{3-m}.  \label{D3}
\end{equation}%
Then we integrate the Eq. (\ref{Hubble alaw}) and follow the guideline in
Appendix (B) to obtain the $\chi ^{2}$ formula to do numerical fitting.
\begin{figure}[tbh]
\includegraphics[width=3.2in,height=3.3in]{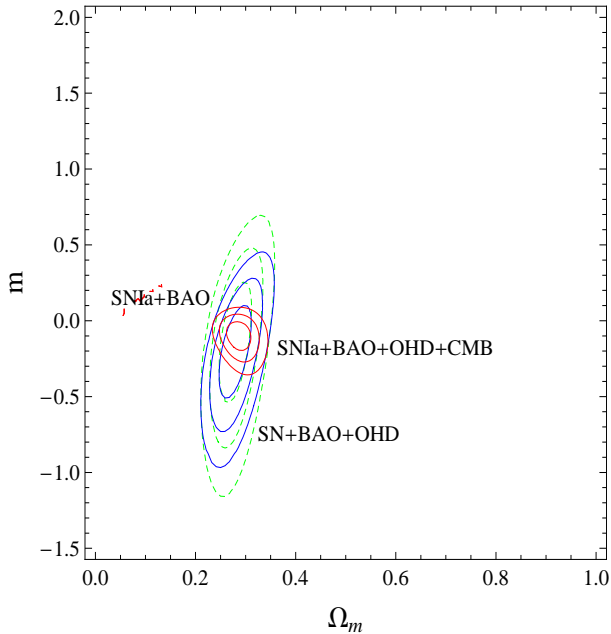}
\caption{Contour map for the parameter $m$ versus $\Omega _{m0}$ in the $a$
power law model. Green dashed lines represent SN+BAO, blue lines represent
SN+BAO+OHD, and red lines represent SN+BAO+OHD+CMB.}
\label{acontour}
\end{figure}
Our result is much tighter than the previous fitting results \cite%
{Silveira97} due to the more precise data we use. Under the combined data
sets SN+BAO+OHD+CMB constraints, the $3\sigma $ values for the power index $%
m $ are
\begin{equation}
m=-0.09_{-0.11-0.20-0.29}^{+0.08+0.12+0.19}.  \label{3sigma}
\end{equation}%
From the above results, we are still not able to rule out the
positive $m$ value within $2\sigma $ confidence level, so we need
more precise data to constrain the VCC models' power index in the
future. However, no matter whether the $m$ is negative or positive,
it is very near zero, suggesting that even though the dark energy is
not a constant, it should evolve very slow. The results for the
best-fit and $1\sigma $ values are shown in Table 1.
\begin{table*}[tbp]
\begin{centering}
%\arraystretch{1.4}
\begin{tabular}{|c|c|c|c|c|}\hline
\multicolumn{2}{|c|}{Models} & SN+BAO & SN+BAO+OHD & SN+BAO+OHD+CMB
\\\hline $a$ power & $m$ & $-0.12_{-0.42}^{+0.40}$ &
$-0.19_{-0.32}^{+0.29}$ & $-0.09_{-0.11}^{+0.08}$ \\ \cline{2-5} law
& $\Omega_{m0}$ & $0.28_{-0.04}^{+0.03}$ & $0.28_{-0.03}^{+0.04}$ &
$0.29_{-0.07}^{+0.03}$ \\ \hline $H$ power & $n$ &
$-0.26_{-0.74}^{+0.67}$ & $-0.76_{-0.74}^{+0.24}$ &
$-0.15_{-0.17}^{+0.14}$ \\\cline{2-5} law & $\Omega_{m0}$ &
$0.28\pm0.04$ & $0.27\pm0.03$ & $0.29\pm0.03$
\\\hline
\end{tabular}%
\caption{Results of the fitting for the two models.} \label{tab1}
\end{centering}
\end{table*}
From our results, the $\Omega _{m0}$ is always around $0.28$, which
is consistent with the \textrm{WMAP} three years results
\cite{Bennett03}.
\subsection{Matter density and deceleration parameter of the $a$ power law
models} 
\begin{figure}[tbh]
\includegraphics[width=3.4in]{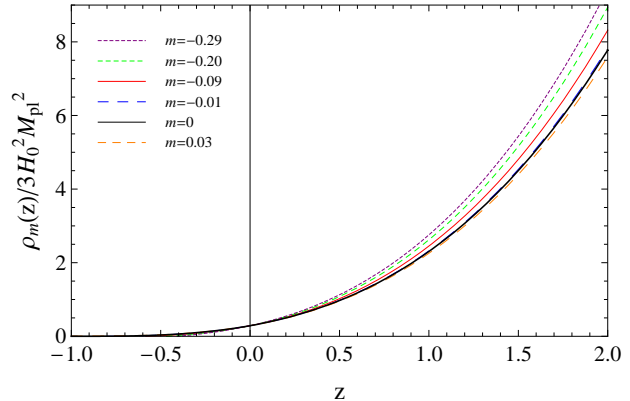}
\caption{Matter density for $a$ power law models. Here we set $\Omega
_{m0}=0.28$.}
\label{arhom}
\end{figure}

\begin{figure}[tbh]
\includegraphics[width=3.4in]{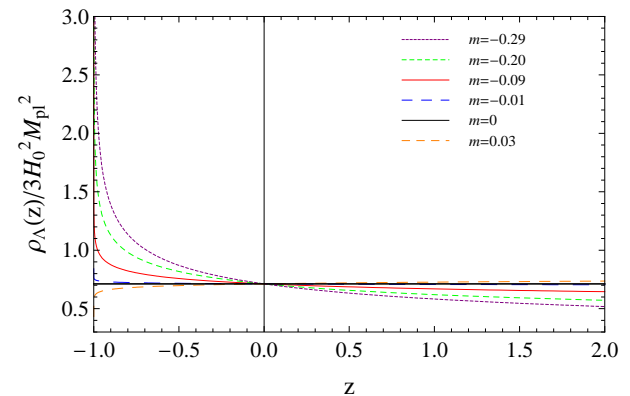}
\caption{Dark energy density for $a$ power law models. Here we set $\Omega
_{m0}=0.28$.}
\label{arhoLamda}
\end{figure}

Having the matter density Eq. (\ref{rho alaw}) and the
confidence region of parameter $m$, we can plot the matter density
and dark energy density as a function of redshift $z$ as FIG.
\ref{arhom} and FIG. \ref{arhoLamda} shows.
We put the curve representing the standard matter density equation in $%
\Lambda $CDM model for comparison.

From FIG. \ref{arhom} and \ref{arhoLamda}, if $m$ is positive, the
dark energy density changes and slower the matter density decreases
as a function of redshift $z$. That means the dark energy "decays"
into dark matter field and makes the matter dilute more slowly with
the cosmic expansion, vice versa. As a result, the dark energy
density will continuously decrease as the Universe evolves. On the
contrary, if $m<0$, the dark matter transforms into dark energy and
the dark matter energy density decreases more sharply than the usual
$(1+z)^{3}$ behavior ($\Lambda $CDM). Then the dark energy increases
its energy density and realizes the "Big Rip" in the future due to
this matter changes, so effectively it resembles the phantom dark
energy (see Eq. (\ref{EoS-alaw})).

It is easy to see this property of index $m$ through the "decay
rate" $\epsilon $ in Ref. \cite{Alcaniz05}. The "decay rate"
$\epsilon $ is defined as the matter density's deviation from the
standard evolution, i.e.,
\begin{equation}
\rho _{m}=\rho _{m0}a^{-3+\epsilon },  \label{9}
\end{equation}%
where $\rho _{m0}$ is the current matter density. In our case, $\epsilon $
is not a constant but a function of redshift $z.$ In addition, it is
straightforward to verify that $m$ is the index to distinguish the sign of $%
\epsilon $, as $m>0$, $\epsilon (z)>0$; vice versa. Thus, the relationship
between index $m$ and the "decay rate" $\epsilon $ represents whether the
dark energy "decays" into dark matter or the inverse. Moreover, it is also
easy to confirm that the value of $\epsilon (z)$ is generally compatible
with the confidence region provided by \cite{Alcaniz05}.

Having obtained some meaning of the index $m$ and its confidence region in
the $a$ power law models, we can directly find the evolution of the
deceleration parameter in this kind of model.
\begin{equation}
q^{(a)}(z)=-\frac{\ddot{a}a}{\dot{a}^{2}}=-\frac{1}{2}\frac{2-m-D_{4}(z)}{%
D_{4}(z)+1},  \label{q for a}
\end{equation}%
where
\begin{equation}
D_{4}(z)=\frac{C_{3}}{C_{2}}(1+z)^{3-m}.  \label{D4}
\end{equation}%
\begin{figure}[tbh]
\includegraphics[width=3.4in]{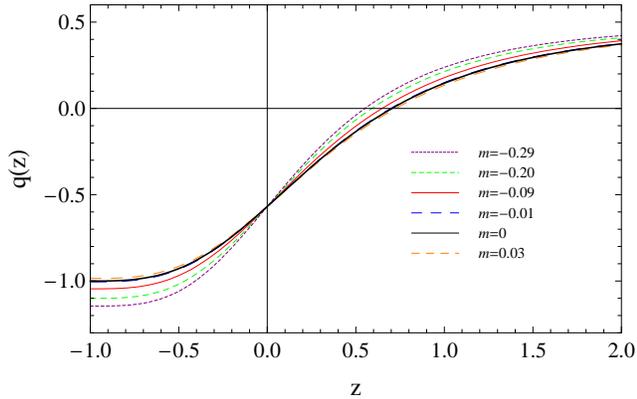}
\caption{Deceleration parameter for $a$ power law models. Here we set $%
\Omega _{m0}=0.28$.}
\label{qalaw}
\end{figure}
From FIG. \ref{qalaw}, we can understand the following characteristics about
the deceleration parameter in the $a$ power law models: First, different $a$
power law behaviors (including $\Lambda $CDM) have common deceleration
parameters at present time $q(0)=\frac{3}{2}\Omega _{m0}-1$. Second, they
have different values of redshift $z$ when the Universe was changing from
deceleration to acceleration, so the transition redshift $z_{T}$ for
different models are determined by the following equation
\begin{equation}
z_{T}=[\frac{(2-m)C_{2}}{C_{3}}]^{\frac{1}{3-m}}-1,  \label{zT for a}
\end{equation}%
which can be determined by the results of the constraints. In addition, we
could calculate the different $z_{T}$ corresponding to $m$ as the best fit, $%
1\sigma $, $2\sigma $ confidence values and the $\Lambda $CDM: $%
z_{T}(m=-0.29)=0.57$, $z_{T}(m=-0.20)=0.61$, $z_{T}(m=-0.09)=0.67$, $%
z_{T}(m=-0.01)=0.72$, $z_{T}(m=0.03)=0.75$. So the larger $m$ is,
the earlier the Universe changes from deceleration to acceleration.
Thus, we obtain this transition redshift $z_{T}$ from the general
$a$ power law form and the results should be applicable for all of
the specific power law behaviors \cite{Alcaniz05}. Third, the more
negative value of $m$ is, i.e., more sharply dark energy density
changes, the faster Universe accelerates. This result is also
compatible with the FIG. 3 in \cite{Alcaniz05}.

Furthermore, we can also use the constraints on the deceleration parameter
and transition redshift to see whether our results are consistent with
relevant constraints\cite{Djorgovski08,Melchiorri07,Alam06}. In Ref. \cite%
{Djorgovski08}, it shows that the best fits for the transition redshift is $%
z_{T}=0.78_{-0.27}^{+0.08}.$ Our results are rather consistent with
this work because this best fits value for $z_{T}$ will lead to the
best fits region for $m$ $[-0.43,0.16]$, and our $3\sigma $ results
for $m$ is just within this region, suggesting that our constraints
on $m$ and $\Omega _{m0}$ are very tight. At the same time, our
results are also very consistent with other relevant constraints on
the "equivalent" redshift when $\rho _{m}(z_{eq})=\rho _{\Lambda
}(z_{eq})$ \cite{Melchiorri07}.

Therefore, in this subsection we conclude that the power index $m$ of the $a$
power law model is not only associated with the dark energy density, but
also a meaningful index to determine whether the dark matter "decays" into
dark energy or the inverse. Moreover, it determines the "decay rate" $%
\epsilon $, i.e., the intensity by which dark matter changes into
dark energy. Meanwhile, it affects the deceleration parameter of the
Universe and the transition redshift $z_{T}$ when the Universe was
changing from deceleration to acceleration.
\section{$H$ power law model and its reconstructed potentials}
In this section, we will discuss another type of VCC models----- $\Lambda $
is associated with Hubble parameter $H$----- which is an important type
presented in Ref. \cite{Overduin98}.

In this type of model, the VCC can be written as
\begin{equation}
\Lambda =CH^{n},  \label{Lamda for H}
\end{equation}%
where $C$ is a constant with the dimension of mass $2-n$, $n$ is the only
parameter in this kind of models which needs to be fitted by observational
data. Then, the dark energy density and the Friedmann equation in this model
can be given by
\begin{equation}
\rho _{\Lambda }^{(H)}=\ \Lambda M_{pl}^{2}=CH^{n}M_{pl}^{2},
\label{rho Lamda for H}
\end{equation}%
\begin{equation}
3M_{pl}^{2}H^{2}=\rho _{m}^{(H)}+\rho _{\Lambda }^{(H)}.  \label{friedmann1}
\end{equation}%
The amplitude $C$ is determined by the current value of matter density and
the Hubble constant
\begin{equation}
C=3H_{0}^{2-n}(1-\Omega _{m0}).  \label{C}
\end{equation}%
Then, we consider that the VCC indicates that there is an interaction
between the dark matter and dark energy. Therefore, let's assume that dark
energy and matter exchange pressure through the interaction term $W(z)$ with
\begin{equation}
\dot{\rho}_{m}^{(H)}+3H\rho _{m}^{(H)}=W(z),  \label{Int matter (H)}
\end{equation}%
\begin{equation}
\dot{\rho}_{\Lambda }^{(H)}+3H(\rho _{\Lambda }^{(H)}+p_{\Lambda
}^{(H)})=-W(z),  \label{Int energy (H)}
\end{equation}%
which maintains the total energy conservation equation $\dot{\rho}%
_{tot}+3H(\rho _{tot}+p_{tot})=0$. Since the VCC is the generalization of
the cosmological constant, so it should satisfy $\rho _{\Lambda
}^{(H)}=-p_{\Lambda }^{(H)}$, then Eq. (\ref{Int energy (H)}) leads to
\begin{eqnarray}
W(z) &=&-\dot{\rho}_{\Lambda }^{(H)}  \nonumber \\
&=&G_{1}H^{n}(1+z)H^{^{\prime }}(z),  \label{W}
\end{eqnarray}%
where
\begin{equation}
G_{1}=3nH_{0}^{2-n}(1-\Omega _{m0})M_{pl}^{2}.  \label{G1}
\end{equation}
\subsection{Interacting dark energy and reconstructed potentials in the $H$
power law model} In this subsection, we want to see the potential
that dark energy mimics the VCC. Although the interaction between
dark energy and dark matter might not be directly observable, the
effective potential could encode some information about the
interaction. Thus, we are looking forward to solving the EEoS and
reconstruct the dark energy potentials of the $H$ power law models
from the standard interacting dark energy Eqs. (\ref{Int matter
(H)}) and (\ref{Int energy (H)}).

\begin{figure}[tbh]
\includegraphics[bb=0 0 362
237,width=3.4in,height=2.5in]{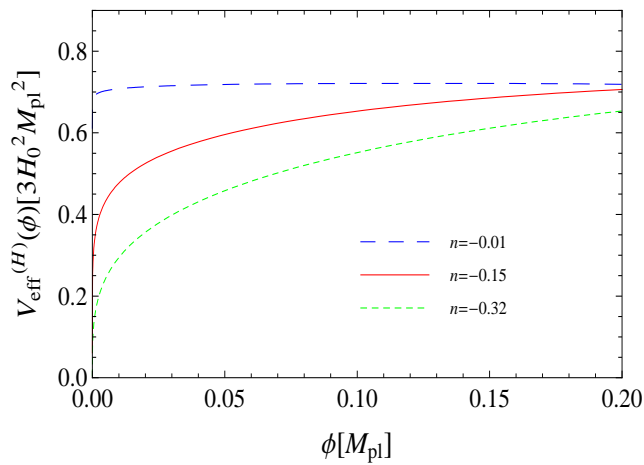}
\caption{Reconstructed potentials for $H$ power law model. Here we set $%
\Omega _{m0}=0.28$.}
\label{Hpotential}
\end{figure}

Transforming Eq. (\ref{Int energy (H)}), we have
\begin{equation}
\dot{\rho}_{\Lambda }^{(H)}+3H(\rho _{\Lambda }^{(H)}+p_{\Lambda }^{(H)}+%
\frac{W(z)}{3H})=0,  \label{energy eq for (H)}
\end{equation}%
from which the interaction changes the effective pressure of this model,
i.e.
\begin{equation}
p_{eff}^{(H)}=p_{\Lambda }^{(H)}+\frac{W(z)}{3H},  \label{p eff (H)}
\end{equation}%
so the EEoS of dark energy is
\begin{equation}
\omega _{eff}^{(H)}=\frac{p_{eff}^{(H)}}{\rho _{\Lambda }^{(H)}}=-1+\frac{n}{%
2}\frac{G_{2}(z)}{G_{2}(z)+G_{3}},  \label{H EEoS}
\end{equation}%
where
\begin{equation}
G_{2}(z)=\Omega _{m0}(1+z)^{3(1-\frac{1}{2}n)},G_{3}=(1-\Omega _{m0}).
\label{G2}
\end{equation}%
These EEoS are functions of redshift $z$, in contrast to the $a$ power law
models, where the EEoS is constant in the Eq. (\ref{EoS-alaw}). It is quite
interesting that the sign of index $n$ also determines whether this dark
energy likes the quintessence or phantom. Moreover, this type of EEoS is
affected by the value of $\Omega _{m0}$, while the EEoS in the $a$ power law
models are not. Since the constraint from the current observational data
suggests that the best-fit for $n$ is negative but cannot rule out the
possibility of positive constant $n$ (see discussion in next section), we
construct the potentials for the two cases. The energy density and pressure
density of the quintessence field for this model are
\begin{equation}
\mp \frac{1}{2}\dot{\phi}^{2}+V_{eff}^{(H)}(\phi )=\rho _{\Lambda }^{(H)},
\label{rho Lamda H}
\end{equation}%
\begin{equation}
\mp \frac{1}{2}\dot{\phi}^{2}-V_{eff}^{(H)}(\phi )=p_{eff}^{(H)}.
\label{p Lamda H}
\end{equation}%
where the upper (lower) sign represents the phantom (quintessence) dark
energy, corresponding to $n<0$ ($n>0$). At the same time, we can obtain the
expressions for dark energy density and pressure through definition (\ref%
{rho Lamda for H}) and the interacting dark energy Eq. (\ref{Int energy (H)}%
).
\begin{equation}
\rho _{\Lambda }^{(H)}=A_{1}[G_{2}(z)+G_{3}]^{\frac{n}{2-n}},
\label{rho Lamda (H)}
\end{equation}%
\begin{eqnarray}
p_{\Lambda }^{(H)} &=&-A_{1}[G_{2}(z)+G_{3}]^{\frac{2(n-1)}{2-n}}  \nonumber
\\
&\times &[(1-\frac{n}{2})G_{2}(z)+G_{3}].  \label{p Lamda (H)}
\end{eqnarray}%
Then, the effective scalar potential can be written as a function of
redshift $z$
\begin{eqnarray}
\tilde{V}_{eff}^{(H)}(z) &=&G_{3}[G_{2}(z)+G_{3}]^{\frac{2(n-1)}{2-n}}
\nonumber \\
&\times &[(1-\frac{n}{4})G_{2}(z)+G_{3}].  \label{H potential}
\end{eqnarray}%
Using the Eqs. (\ref{rho Lamda H}) and (\ref{p Lamda H}), we can obtain the
differential form of scalar field
\begin{equation}
\frac{d\tilde{\phi}}{dz}=\mp \frac{G_{4}(1+z)^{\frac{1}{2}-\frac{3}{4}n}}{%
G_{2}(z)+G_{3}},  \label{H field diff}
\end{equation}%
where
\begin{equation}
G_{4}=(\frac{3}{2}\left\vert n\right\vert )^{\frac{1}{2}}(\Omega
_{m0}(1-\Omega _{m0}))^{\frac{1}{2}} \text{.}  \label{G4}
\end{equation}%
In general, Eq (\ref{H field diff}) could be solved analytically so we
obtain the following field equation
\begin{equation}
\tilde{\phi}(z)=\tilde{\phi}_{0}\mp G_{5}\arctan [G_{6}(1+z)^{\frac{3}{2}(1-%
\frac{1}{2}n)}],  \label{H field anal}
\end{equation}%
where
\begin{equation}
G_{5}=\frac{2}{2-n}(\frac{2}{3}\left\vert n\right\vert )^{\frac{1}{2}%
},G_{6}=(\frac{\Omega _{m0}}{1-\Omega _{m0}})^{\frac{1}{2}},  \label{G5}
\end{equation}%
and the upper(lower) sign applies if $\dot{\phi}>0$ $(\dot{\phi}<0)$. In
fact, the sigh is arbitrarily determined by assumption, as it can be changed
by $\phi \rightarrow -\phi .$ We substitute $(1+z)$ for $\phi $ into Eq. (%
\ref{H potential}) to get the result of potential $\tilde{V}_{eff}^{(H)}(%
\tilde{\phi})$.
\begin{equation}
\tilde{V}_{eff}^{(H)}(\tilde{\phi})=G_{3}^{\frac{2}{2-n}}[1+\tan ^{2}(\frac{(%
\tilde{\phi}-\tilde{\phi}_{0})}{G_{5}})]^{\frac{2(n-1)}{2-n}}.
\label{Veff for H}
\end{equation}%
We give the following three examples of phantom potentials for this kind of
models (see FIG. \ref{Hpotential}).
As is shown in FIG. \ref{Hpotential}, the effective phantom
potentials also have some characteristics: For one thing, they are
all runaway type potentials and even if we change the initial value
$\tilde{\phi}_{0}$, the curves shift horizontally with the whole
shape unchanged. For another thing, the meanings of the potentials
are clear: as the Universe is expanding, the value of $\phi $
becomes large and the field slowly rolls upon the potential, which
makes the EEoS very close to $-1$. At the same time, the dark matter
field gradually "decays" into dark energy, so the dark energy
density increases its energy density as the Universe expands.
\subsection{One specific examples of reconstructed potentials for the $H$
power law models} In Ref. \cite{Overduin98}, there is a list of
proposed $H$ power law models proposed by different authors through
various perspectives. In addition, Ref. \cite{Maia01} gives two
examples of scalar potentials from the point of view of the scalar
field description. In order to show the effectiveness of our
reconstruction, we derive one analytic results of the scalar
potentials using the methods in previous subsection.

$n=2$ is an interesting \cite{Freese87} case of the VCC model and
investigated by many authors \cite{Overduin98,Maia01}. We substitute $n=2$
into Eq. (\ref{H potential}) to obtain
\begin{equation}
\tilde{V}_{eff}^{(H)}(z)=(1-\Omega _{m0})(1-\frac{1}{2}\Omega
_{m0})(1+z)^{3\Omega _{m0}},  \label{Veff (z)}
\end{equation}%
and the field (\ref{H field diff}) could be integrated as
\begin{equation}
\tilde{\phi}(z)=\tilde{\phi}_{0}-3G_{7}\ln (1+z),  \label{phi(z)}
\end{equation}%
where $\tilde{\phi}_{0}$ is the initial value of field $\tilde{\phi}$ and
\begin{equation}
G_{7}=\Omega _{m0}(1-\Omega _{m0}).  \label{G7}
\end{equation}%
Thus, we obtain the potential by substituting $(1+z)$ for field $\tilde{\phi}
$
\begin{equation}
\tilde{V}_{eff}^{(H)}(\tilde{\phi})=(1-\Omega _{m0})(1-\frac{1}{2}\Omega
_{m0})e^{-\alpha (\tilde{\phi}-\tilde{\phi}_{0})},  \label{Veff(phi)}
\end{equation}%
where $\alpha =(\frac{3\Omega _{m0}}{1-\Omega
_{m0}})^{\frac{1}{2}}$. This form is rather consistent with that in
Ref. \cite{Maia01}, which demonstrates the effectiveness of the
reconstructing method in this paper. This potential is one of the
simplest runaway types which represents the particle creation in the
phenomenological VCC models so it could be interpreted as some kind
of "coupled quintessence" \cite{Dalal01}. Meanwhile, it is also easy
to see that all the VCC potentials are associated with the
exponential function, which leads to its runaway behavior,
indicating that they might be easily obtained in supergravity and
unstable D-brane systems \cite{Sen05}.
\subsection{Hubble parameter and results of the constraints on $n$}
From Eqs. (\ref{friedmann1}), (\ref{Int matter (H)}) and (\ref{W}), we can
obtain the differential equation for the Hubble parameter
\begin{equation}
H^{^{\prime }}(z)-\frac{1}{2(1+z)}(3H-K_{1}H^{n-1})=0,  \label{H equation}
\end{equation}%
where
\begin{equation}
K_{1}=3H_{0}^{2-n}(1-\Omega _{m0}).  \label{K1}
\end{equation}%
The Eq. (\ref{H equation}) can be solved analytically%
\begin{equation}
H(z)=H_{0}[G_{2}(z)+G_{3}]^{\frac{1}{2-n}}.  \label{Hubble Hlaw}
\end{equation}%
Then, we follow the procedure in Appendix (A) and (B) to do the numerical
fitting and finally we can find $1\sigma $ results as Table 1 shows. Under
the combined SN+BAO+OHD+CMB constraints, the best fit, $1\sigma ,2\sigma $
and $3\sigma $ values for parameter $n$ is
\begin{equation}
n=-0.15_{-0.17-0.26-0.43}^{+0.14+0.23+0.25}.  \label{3sigma n}
\end{equation}%
\begin{figure}[tbh]
\includegraphics[width=3.2in,height=3.2in]{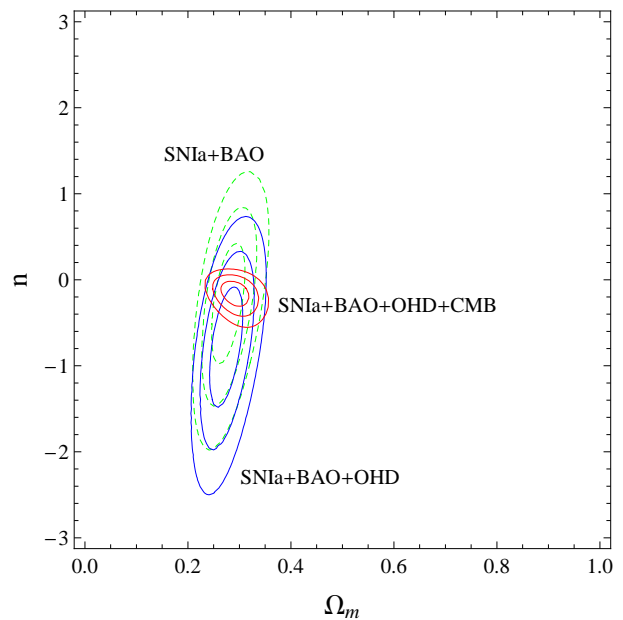}
\caption{Contour map for the parameter $n$ versus $\Omega _{m0}$ in the $H$
power law model. Green lines represent SN+BAO, blue lines represent
SN+BAO+OHD, and red lines represent SN+BAO+OHD+CMB.}
\label{Hcontour}
\end{figure}
From the result (\ref{3sigma n}), no matter whether $n$ is negative
or positive, it always very near $0$, indicating the slow evolution
of dark energy. We could plot the EEoS (\ref{H EEoS}) as a function
of redshift $z$ and compare them with other models \cite{Nesseris04}
and observational results \cite{Nesseris07}.
\begin{figure}[tbh]
\includegraphics[width=3.4in]{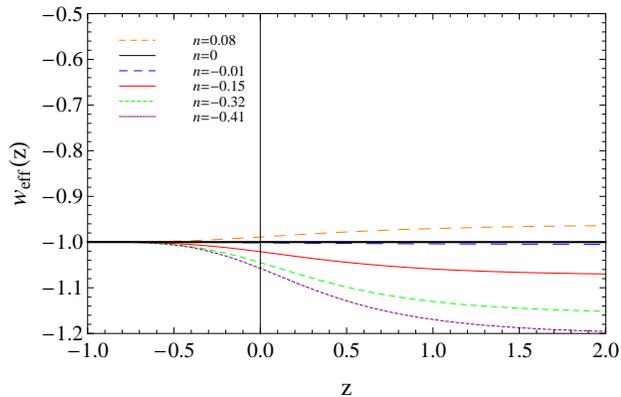}
\caption{The redshift dependence of the EEoS for $H$ power law models. Here
we set $\Omega _{m0}=0.28$.}
\label{HEEoS}
\end{figure}
This EEoS has three major properties: For one thing, the confidence region
of this type of models is mildly consistent with the results which were
obtained by using CMB and Clusters data \cite{Nesseris07}, indicating that
it is a competitive model waiting for the examination by future
observations. For another thing, within $1\sigma $ confidence region, the
EEoS is slightly less than $-1$, which implies that it resembles the phantom
field. In contrast, within $2\sigma $ region it is possible that the EEoS is
greater than $-1$, so we cannot rule out the possibility that the dark
energy is quintessence like. Thus, this type of models could really
represent a large kind of dark energy models phenomenologically. Further
more, if $z$ become larger, all of the EEoS in these models have their own
asymptotically constant value, which is rather similar to that of the
"quiessence model" \cite{Sahni03}. The constant EEoS means that the
proportion of kinetic energy to potential energy is constant. Thus, the
whole dark energy density increases or decreases, suggesting that the VCC
models corresponds to a dissipative system of dark energy \cite%
{Strominger89,Ozer86,Dalal01}. When redshift $z$ approaches $-1$,
all of the EEoS $w_{eff}(z)$ approach $-1$, indicating the Universe
will enter the de-Sitter phase in the future.
\subsection{Matter density, Hubble parameter and deceleration parameter of
the $H$ power law models} From the Hubble parameter Eq. (\ref{H
equation}), we can obtain the matter density
\begin{equation}
\rho _{m}^{(H)}(z;n)=3H_{0}^{2}M_{pl}^{2}G_{2}(z)[G_{2}(z)+G_{3}]^{\frac{n}{%
2-n}}\text{.}  \label{rhom for H}
\end{equation}%
Then, we plot the dark matter density and Hubble parameter of the $H$ power
law model. In order to compare with $\Lambda $CDM model, we also plot $n=0$
curve in one graph.
\begin{figure}[tbh]
\includegraphics[width=3.4in]{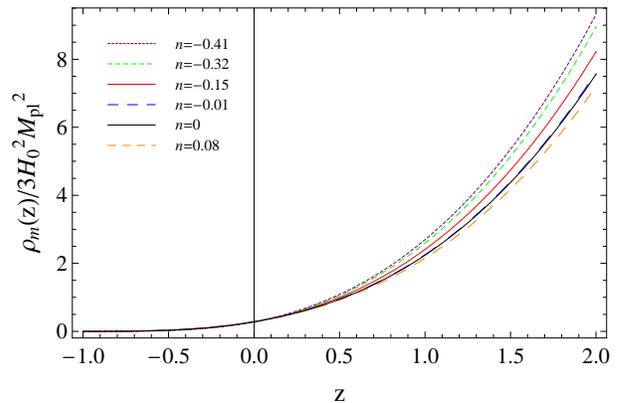}
\caption{Matter density for $H$ power law models. Here we set $\Omega
_{m0}=0.28$.}
\label{Hrhom}
\end{figure}
Thus, FIG. \ref{Hrhom} helps us to analyze the properties of the matter
density in this model: For one thing, we can see that for the positive $n$,
the dark energy density decreases and the matter density dilutes more slowly
as the Universe evolves, vice versa. On the contrary, $n<0$ represents the
dark matter changes into dark energy because the dark energy density
increases as time evolves and the corresponding curve (for example, the
curve in the graph $n<0$) decreases more sharply than the standard $%
(1+z)^{3} $ behavior ($\Lambda $CDM). As a result, the parameter $n$ is not
only a power index of the $H$ power law models, but also an important
signature to distinguish whether dark energy "decays" into dark matter or
the inverse process, just as the index $m$ in the $a$ power law case.
Further more, comparing FIG. \ref{Hrhom} and FIG. \ref{arhom}, we can
discover that the two graphs are very similar to each other, which means
that the two types of VCC models----- the $a$ power law and the $H$ power
law models----- really share some common features if the parameters are all
constrained by observational data. This could be understood as follows: both
the scale factor $a(t)$ and the Hubble parameter $H(z)$ describe the
evolution of the Universe; if the Universe is expanding canonically, $a(t)$
will increase while the $H(z)$ will decrease, so the difference between the
two types might only lie in the sign of the power index.

However, if we plot the Hubble parameter by selecting some ideal value of $n$%
, we could see that they indicate the different fates of the Universe.
\begin{figure}[tbh]
\includegraphics[width=3.4in]{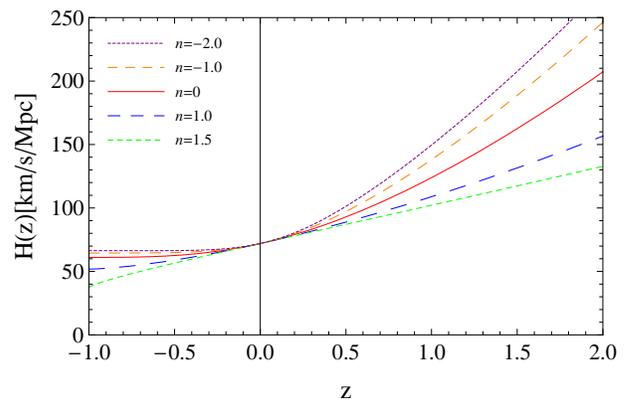}
\caption{Hubble parameter for $H$ power law models. Here we set $\Omega
_{m0}=0.28$ and $H(0)=72\text{km/s/Mpc}$.}
\label{Hparameter}
\end{figure}
We see from Fig. (\ref{Hparameter}) that for any selected value of $n$, the
Hubble parameter cannot deverge, so the $H$ power law model does not
indicate the "Big Rip" phase in the future. This property could be
understood from Fig. (\ref{HEEoS}), because whatever the current value of $%
w_{eff}$ is, they will all tend to be $-1,$ so the Universe will enter a
de-Sitter phase in the future. This property is rather different from $a$
power law model since $w_{eff}$ is a constant in that model, so the dark
energy will be always a phantom if $m<0$, which surely results in a phantom
Universe with the "Big Rip" phase in the future (see Fig. (\ref{arhoLamda})).

Having obtained the matter density, we can derive the deceleration parameter
in this model
\begin{equation}
q^{(H)}(z)=-\frac{1}{2}\frac{2-K_{2}(z)}{1+K_{2}(z)}\text{,}  \label{q for H}
\end{equation}%
where%
\begin{equation}
K_{2}(z)=\frac{\Omega _{m0}}{1-\Omega _{m0}}(1+z)^{3(1-\frac{1}{2}n)}\text{,}
\label{K2}
\end{equation}%
Then, we can plot this deceleration parameter and also compare the curves
with that of $\Lambda $CDM.
\begin{figure}[tbh]
\includegraphics[width=3.3in]{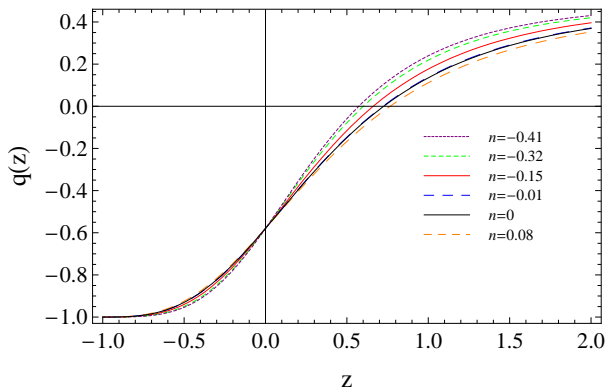}
\caption{Deceleration parameter for $H$ power law models. Here we set $%
\Omega _{m0}=0.28$.}
\label{qHlaw}
\end{figure}
From FIG. (\ref{qHlaw}), we could analyze the properties of the deceleration
parameter in this $H$ power law model. For one thing, just as the $a$ power
law case, all of the $H$ power law models share the same deceleration
parameter at present time, which is $q(0)=\frac{3}{2}\Omega _{m0}-1$, but
they also share the same deceleration parameter values in the future ($%
a\rightarrow \infty $) as $q(z=-1)=-1$, which is different from that of the $%
a$ power law model. For another thing, transition redshift $z_{T}$ varies
differently according to the different curves, which indicates that
different values of $n$ affect the expansion of the Universe distinctively.
The transition redshift $z_{T}$ for different models are determined by the
following equation
\begin{equation}
z_{T}=[\frac{2(1-\Omega _{m0})}{\Omega _{m0}}]^{\frac{2}{3(2-n)}}-1.
\label{zT for H}
\end{equation}%
We could calculate its value with respect to different index $n$ numerically
(see Table 2). It is also easy to see that the value of $z_{T}$ within $%
1\sigma $ confidence region is generally compatible with the result
in Ref. \cite{Alcaniz05}. In addition, from Fig. (\ref{qHlaw}), the
transition redshift is in the range $[0.57,0.77]$, which is just in
the best fits
region of transition redshift in Ref. \cite{Djorgovski08} and \cite%
{Melchiorri07}, indicating that our numerical constraints on the parameters $%
n$ has been already very tight compared with relevant work
\cite{Djorgovski08,Melchiorri07,Alam06}. Moreover, different values
of $n$ correspond to different curves with distinctive shapes. For
one thing, within $1\sigma $ region, $n$ is definitely negative
(such as $n=-0.15$), this makes the dark energy density increases
and dark matter "decays" into dark energy as time evolves. The more
negative the index $n$ is, the more quickly the dark energy density
$\rho _{\Lambda }^{(H)}$ increases and the faster the Universe
accelerates. For another thing, within $2\sigma $ confidence region,
there is a certain probability that the index $n$ is positive, which
indicates that the density of dark energy is decreasing, so the
acceleration is relatively small compared to the negative value of
$n$.
\begin{table*}[tbp]
\begin{widetext}
\begin{center}
\begin{tabular}{c|lllllllll}
\hline\hline $n$ & $-0.41(2\sigma )$ & $-0.32(1\sigma )$ &
$-0.15$(the best fit) & $-0.01(1\sigma )$ & $0$($\Lambda $CDM) &
$0.08$($2\sigma $) & & &  \\ \hline
$z_{T}$ & $0.57$ & $0.60$ & $0.66$ & $0.72$ & $0.73$ & $0.77$ &  &  &  \\
\hline\hline
\end{tabular}%
\end{center}
\caption{transition redshift $z_{T}$ in $H$ power law model.}
\label{tab2}
\end{widetext}
\end{table*}
To conclude this subsection, we derive the evolution of the matter
density, Hubble parameter and the deceleration parameter in the $H$
power law models. We note that the index $n$ is just like the index
$m$ in the $a$ power law models, which not only reflects whether the
dark energy "decays" into dark matter, but also affects the
acceleration of the Universe----- the deceleration parameter $q(z)$
and transition redshift $z_{T}$.
\section{Concluding Remarks}
In this paper, we develop a method to reconstruct potentials in the
VCC models directly from the definition of the energy density and
pressure of the scalar field. We also give one example of the
reconstruction for the $H$ power law models. First, these potentials
have some relationship with the exponential function, which is
expected in supersymmetry theory and unstable D-brane system in
superstring theory \cite{Sen05}. Second, as the Universe is
expanding, the value of $\phi $ becomes large and the field slowly
rolls upon the potential. At the same time, the dark matter field
gradually "decays" into dark energy, so the matter density dilutes
more sharply than the standard $(1+z)^{3}$ behavior. It is worth
noticing that the reconstruction equations presented here are not
limited to searching for the scalar field description of such
phenomenological VCC models. Generally, it could give people the
possibility to find the scalar field versions of other
phenomenological models and even quantum gravity models such as
holographic models \cite{MiaoLi04} and vacuum fluctuation model
\cite{Djorgovski06}.

We also investigate constraints on the VCC models----- the $a$ power law and
the $H$ power law models----- from current combined cosmological
observations, using high quality supernovae, baryon acoustic oscillation
from SDSS, observational $\text{H(z)}$ data derived from Gemini Deep Deep
Survey (GDDS) and the CMB shift parameter from \textrm{WMAP} three years
result. We consider a spatially flat FRW Universe with matter component and
VCC component. For the VCC models, such as the $a$ power law and the $H$
power law models, the power indices $m$ and $n$ play a very significant role
in determining the evolutionary behavior of the space-time as well as the
ultimate fate of the Universe. According to the combined constraints, the
best fit, $1\sigma $, $2\sigma $ and $3\sigma $ values for the indices $m$
and $n$ are $m=-0.09_{-0.11-0.20-0.29}^{+0.08+0.12+0.19}$, $%
n=-0.15_{-0.17-0.26-0.43}^{+0.14+0.23+0.25}$ respectively. These
results cannot fix the value of $m$ and$n$, but at least indicate
even though the dark energy may change with time, it will evolve
very slow since the value of $m$ and$n$ are always very near zero.
The indices of VCC models suggest the interaction between dark
energy and dark matter: First, $m>0$ and $n>0$ represent that the
dark energy "decays" into dark matter, while $m<0$ and $n<0$
represent the inverse process. The more negative the indices $m$ and
$n$ are, the faster such transitions happen. Second, the indices $m$
and $n$ affect the deceleration parameter and the transition
redshift $z_{T}$. The more negative the indices $m$ and $n$ are, the
more portion dark energy takes in the whole Universe budget in the
future due to the matter "decays"; thus, faster the Universe will be
accelerating. Also, the more negative the indices $m$ and $n$ are,
the less portion they took into the whole Universe budget in the
past and the smaller value of transition redshift $z_{T}$ is.
Therefore, the indices $m$ and $n$ are the important signatures to
judge whether the Universe accelerates more drastically than the
$\Lambda $CDM model. Third, the indices $m$ and $n$ are also the
essential indicators to understand the properties of VCC. For one
thing, the best fit values for $m$ and $n$ suggest that the EEoS of
dark
energy are real numbers or functions of redshift $z$ at the region $%
[-1.5,-1] $, indicating that the dynamic scalar fields of VCC are
phantom-like. For another thing, there are still some probabilities for the $%
a$ power law and the $H$ power law models that the dynamic scalar field of
VCC is quintessence-like, so the VCC models are the phenomenal models
representing a variety of other dynamic dark energy models. Moreover, the
EEoS of the $a$ power law models are constant, while those of the $H$ power
law models are functions of redshift $z$ but have the asymptotic values when
redshift $z$ becomes large, so at the early stage, the VCC models have some
properties of the quiessence model. The different fate indicated by the two
models are, if redshift $z$ tends to be $-1$, the EEoS in $H$ power tends to
be $-1,$ so the Universe enter a de-Sitter phase in the future. However, the
EEoS of $a$ power law model is always a negative constant, indicating the
Universe will enter the "Big Rip" phase in the future.

The cosmological constant problem is still one of the serious
problem that puzzles the physical world and we are still far away to
go to its nature. Thus, we expect that a more sophisticated combined
analysis of various observations will be capable of determining the
indices value of VCC models and revealing more properties of the VCC
dark energy models.
\section*{Acknowledgments}
The author would like to thank Hui Li, Rong-Gen Cai, Shi Qi, Tan Lu, Yan
Gong and Yong-Shi Wu for helpful discussion. He also thanks Guo-Xing Ju, Hao
Yin, Lei Zhang, Nan Zhao and Tie-Yan Si for helpful direction of computer
program and Hao Wei, Hong-Sheng Zhang, Hui Li, Li-Ming Cao, Xing Wu for
revision and Yan Gong for suggesting high quality supernovae data sets. This
work was supported partially by grants from NSFC, China (No. 10325525, No.
90403029 and No. 10525060) and a grant from the Chinese Academy of Sciences.
\appendix
\section{Some results of integrals}
We define a dimensionless function
\begin{equation}
E(z)\equiv H(z)/H_{0},  \label{E(z)}
\end{equation}%
and integral
\begin{equation}
I(z)\equiv \int_{0}^{z}\frac{dz^{^{\prime }}}{E(z^{^{\prime }})}.
\label{Integral}
\end{equation}%
For the $a$ power law model, using Eq. (\ref{Hubble alaw}), we have%
\begin{eqnarray}
I_{1}(z) &=&\frac{2}{(2-m)D_{3}^{\frac{1}{2}}}[(1+z)^{-\frac{m}{2}+1}
\nonumber \\
&&\times F_{1}(\alpha _{1},\beta _{1},\gamma _{1},\delta _{1}(1+z)^{3-m})
\nonumber \\
&&-F_{1}(\alpha _{1},\beta _{1},\gamma _{1},\delta _{1})],
\label{Integral alaw}
\end{eqnarray}%
where
\begin{eqnarray}
\alpha _{1} &=&\frac{2-m}{2(3-m)},\text{ }\beta _{1}=\frac{1}{2},
\label{alpha} \\
\gamma _{1} &=&\frac{8-3m}{2(3-m)},\text{ }\delta _{1}=-\frac{C_{3}}{C_{2}},
\label{gamma}
\end{eqnarray}%
and $F_{1}(\alpha ,\beta ,\gamma ,x)$ represents the hypergeometric function.

For the $H$ power law model, using Eq. (\ref{Hubble Hlaw}), we obtain
\begin{eqnarray}
I_{2}(z) &=&H_{0}^{-1}(1+z)\frac{2}{(2-m)D_{3}^{\frac{1}{2}}}[(1+z)^{-\frac{m%
}{2}+1}  \nonumber \\
&&\times F_{1}(\alpha _{1},\beta _{1},\gamma _{1},\delta _{1}(1+z)^{3-m})
\nonumber \\
&&-F_{1}(\alpha _{1},\beta _{1},\gamma _{1},\delta _{1})].
\label{Integral Hlaw}
\end{eqnarray}
\section{Data Analysis for the Numerical Fitting}
We utilize several data sets to constrain the parameters of the two
power law model. The free parameters in these two models are power
law index $m$ (or $n$), current value of fractional energy density
of dark matter $\Omega _{m0}$, and the current value of $h$
($h=H_{0}/100/km\cdot s^{-1}\cdot Mpc^{-1}$). However, since the
current value of Hubble parameter $h$ is always around $0.70$ which
is determined by current supernovae constraints so we marginalize it
and plot the contour maps of power index $m$ and $n$ versus $\Omega
_{m0}.$ Our data sets include $307$ high quality "Union" SN Ia data,
baryon acoustic oscillation measurement from the Sloan Digital Sky Survey, $%
9 $ observational $H(z)$ data and the shift parameter from
\textrm{WMAP} three years results.
\subsection{Selected high quality SN Ia data set}
The first standard candle we use is the type Ia supernovae (SNe Ia),
which is published by Supernova Cosmology Project (SCP) team
recently \cite{Kowalski08}. This data set contains 307 selected SNe
Ia that includes several current widely used SNe Ia data, such as
Hubble Space Telescope (HST) \cite{Riess04,Riess06}, SuperNova
Legacy Survey (SNLS) \cite{Astier06,Nesseris06} and the Equation of
State: SupErNovae trace Cosmic Expansion (ESSENCE) \cite{Wood06}.
The likelihood function can be determined by $\chi ^{2}$ statistics,
for the type Ia supernovae
\begin{equation}
\chi _{\mathrm{SN}}^{2}=\sum_{i=1}^{182}\frac{(\mu _{th}(\mathrm{parameters}%
;z_{i})-\mu _{\exp }^{(i)})^{2}}{\sigma _{i}^{\ast 2}},  \label{chiSN}
\end{equation}%
where
\begin{equation}
\mu _{th}(\mathrm{parameters};z)=5\log d_{L}(z)+25,  \label{miuSN}
\end{equation}%
where $d_{L}$ is the luminosity distance which is determined by Eq. (\ref%
{Integral})
\begin{equation}
d_{L}(z)=(1+z)\int_{0}^{z}\frac{dz^{^{\prime }}}{H(z^{^{\prime }})}%
=H_{0}^{-1}(1+z)I(z).  \label{Luminosity distance}
\end{equation}
\subsection{Baryon Acoustic Oscillation measurement from SDSS}
In the large-scale clustering of galaxies, the baryon acoustic oscillation
signatures could be seen as a standard ruler providing the other important
way to constrain the expansion history of the Universe. We use the
measurement of the BAO peak from a spectroscopic sample of 46,748 luminous
red galaxies (LRGs) observations of SDSS to test cosmology \cite%
{Eisenstein05}, which gives the value of $A=0.469(n_{s}/0.98)^{-0.35}\pm
0.017$ at $z_{\mathrm{BAO}}=0.35$ where $n_{s}=0.95$ \cite{Spergel06}. The
expression of $A$ can be written as
\begin{eqnarray}
A &=&\frac{\sqrt{\Omega _{m0}}}{(H(z_{\mathrm{BAO}})/H_{0})^{\frac{1}{3}}}[%
\frac{1}{z_{\mathrm{BAO}}}\int_{0}^{z_{\mathrm{BAO}}}\frac{dz^{^{\prime }}}{%
H(z^{^{\prime }})/H_{0}}]^{\frac{2}{3}}  \nonumber \\
&=&\frac{\sqrt{\Omega _{m0}}}{E(z_{\mathrm{BAO}})^{\frac{1}{3}}}[\frac{I_{%
\mathrm{BAO}}}{z_{\mathrm{BAO}}}]^{\frac{2}{3}}  \label{BAO}
\end{eqnarray}%
and the $\chi _{BAO}^{2}$ is
\begin{equation}
\chi _{BAO}^{2}=\Big(\frac{A-0.469(n_{s}/0.98)^{-0.35}}{0.017}\Big)^{2}.
\label{chiBAO}
\end{equation}
\subsection{Observational H(z) Data (OHD)}
By using the differential ages of passively evolving galaxies determined
from the Gemini Deep Deep Survey (GDDS) and archival data \cite{GDDS}, Simon
et al. determined H(z) in the range $0<z<1.8$ \cite{Simon04}.The 9
observational $H(z)$ pieces of data could be obtained from \cite%
{Simon04,Wei06} and they have been have been used to constrain the dark
energy potential and equation of state \cite{Wei06}. The $\chi ^{2}$
statistics for these $H(z)$ data is
\begin{equation}
\chi _{\mathrm{OHD}}^{2}=\sum_{i=1}^{9}\frac{(H(\mathrm{parameters}%
;z_{i})-H_{i})^{2}}{\sigma _{i}^{\ast 2}}.  \label{chiOHD}
\end{equation}
\subsection{CMB Data from \textrm{WMAP} three years results}
The CMB shift parameter may provide an effective way to constrain the
parameters of dark energy models since it has the very large redshift
distribution and be able to constrain the evolution of dark energy very
well. The shift parameter $R$ which is derived from the CMB data takes the
form as
\begin{eqnarray}
R &=&\sqrt{\Omega _{m0}}\int_{0}^{z_{\mathrm{CMB}}}\frac{dz^{\prime }}{%
H(z^{\prime })/H_{0}}  \nonumber \\
&=&\sqrt{\Omega _{m0}}I(z_{\mathrm{CMB}})  \label{shift parameter}
\end{eqnarray}%
The \textrm{WMAP}3 data gives $R=1.70\pm 0.03$ \cite{wang07}, thus we have
\begin{equation}
\chi _{\mathrm{CMB}}^{2}=\Big(\frac{R-1.70}{0.03}\Big)^{2}.  \label{chiCMB}
\end{equation}%
To break the degeneracy and explore the power and differences of the
constraints for these data sets, we use them in several combinations to
perform our fitting: \textrm{SN + BAO}, \textrm{SN + BAO + OHD}, and \textrm{%
SN + BAO + BAO + CMB}.

\end{document}